\documentclass[aps,showpacs,pra,twocolumn]{revtex4-1}

\usepackage[normalem]{ulem}
\usepackage{exscale}
\usepackage{graphicx}
\usepackage{amsmath}
\usepackage{latexsym}
\usepackage{amsfonts}
\usepackage{amssymb}
\usepackage{times}
\usepackage[T1]{fontenc}
\usepackage{lipsum}
\usepackage{amsthm}

\begin{document}

\title{Comment on 'Reasonable fermionic quantum information theories require relativity'}

\author{Markus Johansson} 
\affiliation{ICFO-Institut de Ciencies Fotoniques, Barcelona Institute of Technology, 08860 Castelldefels (Barcelona), Spain}

\begin{abstract}

In [N. Friis, New J. Phys. {\bf 18}, 033014 (2016)] the non-relativistic description of fermions is considered and in particular the role of the parity superselection rule in relation to the characterization of entanglement.
An argument based on the spin-statistics connection is presented as a physical motivation for the parity superselection rule. Since the spin statistics connection was derived in the context of relativistic quantum mechanics it is argued that the inclusion of the parity superselection rule is motivated by Lorentz invariance.
Based on this it is further argued that fermionic Quantum Information theory and the theory of special relativity are conceptually inseparable. In this comment a different, and well known, motivation for the parity superselection rule is given that does not rely on arguments from relativistic quantum mechanics, but instead uses the assumptions that the laws of physics are the same for different observers and the no-signalling principle.

\end{abstract}

\maketitle

Fermions exist naturally in the context of relativistic quantum mechanics as solutions to the relativistic Dirac equation \cite{dirac}.
It therefore makes sense to understand the properties of fermionic systems in a relativistic context. However, non-relativistic descriptions of fermions are commonly used in circumstances where relativistic effects are negligible. 

A necessary property of the description of fermions is that there cannot exist a coherent superposition of an even and odd number of fermions. This restriction is called the {\it parity superselection rule} \cite{wightman,wigner}. Historically it was first formalised by Wick, Wightman, and Wigner \cite{wightman} and is necessary for the consistency of relativistic quantum theory, i.e., it is a consequence of the postulates of quantum mechanics in conjunction with the principles of special relativity (See e.g. Ref. \cite{Weinberg}). One way to motivate the parity superselection rule (SSR) is via the spin-statistics connection \cite{fierz,pauli,schwinger}. The spin statistics connection makes use of Lorentz invariance and implies that fermions have half-integer spin. The transformation properties of spinors then imply that a coherent superposition of an even and an odd number of fermions is not invariant under a $2\pi$ spatial rotation of the system \cite{wigner}.

From the above argument one may be led to believe that the Lorentz invariance of special relativity is a crucial assumption that is needed to motivate the parity SSR. However, in a non-relativistic context the parity SSR can also be derived from the assumption that the laws of physics are the same to all different observers and the microcausality principle. 
In the non-relativistic setting the microcausality principle (See e.g. Refs. \cite{bogolubov,greiner}) states that simultaneous physical operations in disjoint regions of space commute.
If the Hilbert space of the theory is separable, i.e., if it has a countable orthonormal basis, the microcausality principle is equivalent to the no-signalling principle \cite{luder}.
The no-signalling principle states that an observer can not instantly change the outcome statistics of an experiment performed by another distant observer.

Separability of the Hilbert space of a quantum system is a part of the postulates of quantum mechanics as formulated by Dirac \cite{dirac} and von Neumann \cite{neumann}. In particular, the Hilbert space of $n$ fermionic modes, considered in Ref. \cite{friis}, is separable for any natural number $n$. Therefore, no-signalling and microcausality are equivalent in this context.

The equivalence of the microcausality principle and the no-signalling principle, for separable Hilbert spaces, can be understood in the following way. If an observable $O_A$ of an observer A does not commute with an observable $O_B$ of observer B it follows that $O_A$ and $O_B$ do not have a common basis of eigenstates. Thus, at least one eigenspace $S$ of $O_A$ has no basis that is a set of eigenstates of $O_B$. Therefore, there exist an eigenstate $|\psi_A\rangle\in{S}$ which has a nonzero projection onto an eigenvector $|\psi_B\rangle$ of $O_B$ which does not belong to $S$.
If A and B prepare the state of their shared system to be $|\psi_A\rangle$, it follows that A would find the measurement outcome corresponding to $S$ with certainty upon performing the measurement.
However, if B performs the measurement corresponding to $O_B$ the state of the system will be reduced to an eigenstate of $O_B$ and by assumption there is thus a nonzero probability that it is reduced to $|\psi_B\rangle$ which is not in $S$.
Therefore, if B performs the measurement before A the measurement statistics of A changes and A does no longer see the outcome corresponding to $S$ with certainty. Thus, B can instantly change the outcome statistics of A by performing a measurement.
If on the other hand $[O_A,O_B]=0$ it follows that $O_A$ and $O_B$ have a common eigenbasis and therefore every eigenspace of $O_A$ has a basis of eigenstates of $O_B$. Then B cannot change the outcome statistics of A.

To understand the role of the no-signalling principle and the principle that the laws of physics are the same to different observers in the non-relativistic description of fermions we first review the defining properties of non-relativistic fermionic quantum theory.
In such a theory all operators can be described as algebraic combinations of the creation and annihilation operators, $a_i^{\dagger}$ and $a_i$, of fermions in the different modes indexed by $i$. These operators satisfy anti-commutation relations given by

\begin{eqnarray}\label{anti}
\{a_i,a_j\}=0,\phantom{u}\{a_i^{\dagger},a_j^{\dagger}\}=0,\phantom{u}\{a_i,a_j^{\dagger}\}=\delta_{ij}
\end{eqnarray}

Any operator $O^m$ on a set of $m$ fermionic modes can be divided into an odd part $O_{odd}^m$ and and an even part $O_{even}^m$. The odd part contains only monomials of an odd number of creation and annihilation operators, e.g. $a_i$, $a_i^{\dagger}$, and $a_ia_ja_k^{\dagger}$.
Likewise, the even part contains only monomials of an even number of creation and annihilation operators, e.g. $a_ia_j$, $a_i^{\dagger}a_j^{\dagger}$, and $a_ia_ja_k^{\dagger}a_l$.
An odd monomial can change the number of fermions by an odd number and an even monomial by an even number. Thus, coherent superpositions of even and odd numbers of fermions can be created by operators with both even and odd part.
Therefore, if the set of operators is restricted to only even operators the parity SSR is respected.

Now consider two non-overlapping sets of modes.
Any odd part of an operator on one of the two sets of modes anti-commutes with any odd part of any operator on the other set of modes. This follows immediately from the anti-commutation relations of the individual creation and annihilation operators.
An even operator on one set of modes on the other hand commutes with any operator on the other set.
Using this we can now see how the no-signalling principle and the principle that the laws of physics are the same for all observers implies the parity SSR.

Consider two observers that each have access to a set of modes and that their respective sets of modes do not overlap. Then the no-signalling principle implies that the operations that can be performed by one of the observers must commute with those that can be performed by the other observer. 
Since any odd operator of one observer anti-commutes with any odd operator by the other observer it follows that at least one of the two observers must be unable to perform any operations with an odd part.
If we further assume that observers in different locations are subject to the same laws of physics, and therefore are subject to the same limitations on the operations they are able perform, it follows that both observers must be unable to perform any operation with an odd part. Thus, the parity SSR is implied by the two above assumptions.

To better understand the consequences of not imposing the parity SSR we consider how an observer B can change the reduced state of another distant observer A in the unrestricted non-relativistic fermionic formalism.
As described above, if the parity SSR is not imposed the operations performed by distant observers do not necessarily commute. Consider two modes AB, and let ${{\boldsymbol{ \mathcal O }}}_{A}$ be the operator subalgebra generated by the creation and annihilation operators of a given mode A, as in Sect. 3 of Ref. \cite{friis}. The reduced state $\rho_A$ on mode A can be defined with respect to ${{\boldsymbol{ \mathcal O }}}_{A}$ by demanding that the expectation value of any operator in ${{\boldsymbol{ \mathcal O }}}_{A}$ yield the same result for the global state ${\rho }_{{AB}}$, and for $\rho_A$ (See Eq. 4 in Ref. \cite{friis}).

If coherences are allowed between even and odd numbers of fermions we can consider the pure state $\frac{1}{\sqrt{2}}(a^{\dagger}|0\rangle+|0\rangle)$, where $|0\rangle$ is the vacuum state, and the hermitian unitary operator $a^{\dagger}+a$. The expectation value of $a^{\dagger}+a$ for this state is 1. If a unitary operation $b^{\dagger}+b$ is applied to mode B the state changes to $\frac{1}{\sqrt{2}}(b^{\dagger}a^{\dagger}|0\rangle+b^{\dagger}|0\rangle)$. The expectation value of $a^{\dagger}+a$ for this new state is -1. Thus, a unitary operation on B can change the expectation value of a hermitian operator on A. This implies that the reduced state $\rho_A$, as defined in Eq. 4 of Ref. \cite{friis}, has also changed. Moreover, if an operation on mode B can change the reduced state of A, so can operations on any any other mode.
The reduced state of a given observer is therefore in general not invariant under operations performed by other observers, regardless of their distance.

The violation of no-signalling inherent in the unrestricted fermionic formalism is a direct consequence of the anti-commutation relations \eqref{anti}. Physically, these relations can be understood as encoding the braiding properties of fermions.
When two fermions are exchanged through a braiding operation the state vector acquires a phase factor -1. This is true for any pair of fermions regardless of their spatial separation since braiding operations can exchange fermions that are arbitrarily far apart. This property of braiding is non-local in the sense that it involves arbitrary spatial separation, but no-signalling is not violated since a braiding operation requires the particles to be physically transported from one location to the other. Such a particle transport is represented by an annihilation of the particle in one mode and creation in another mode, which is an even operator, e.g. $a_j^{\dagger}a_i$.
Thus, the violation of no-signalling in the unrestricted fermionic formalism is an artefact of the braiding properties when the parity SSR is not respected.

\subsection*{Discussion}

The purpose of this comment is to give an account of the commonly used motivation for the parity superselection rule based on the no-signalling principle, or microcausality, as an alternative to the motivation involving Lorentz invariance. While widely used in the context of non-relativistic theories this argument is often not explicitly stated.
In addition, an example to illustrate the physical role of the parity SSR in preventing signalling in the non-relativistic quantum theory of fermions was given.

In Ref. \cite{friis} it is argued that Lorentz invariance is required to motivate the parity SSR, and thus that it cannot be properly motivated within a non-relativistic description of fermions. This argument is based on the assumption that the spin statistics connection is a necessary part of the motivation.
From this Ref. \cite{friis} argues that "any fermionic QI theory {\it must} be seen as (part of) a relativistic QFT." 
It does not consider if the parity SSR can be motivated by non-relativistic physical principles.

Here it is argued that such physical principles do not only exist but are commonly used in non-relativistic theories, including those that do not describe fermions. Moreover, the argument based on these principles relies only on the anti-commutation relations of fermionic operators and does not involve spin.

In conclusion, it is possible to motivate the parity SSR in the non-relativistic context without referral to Lorentz invariance. An argument can be given based on the assumption of no-signalling and the assumption that the laws of physics are the same for different observers. 
Thus, fermionic Quantum Information theory does not need to be seen as part of a relativistic QFT unless we posit that any other non-relativistic quantum theory that satisfies these two principles must be seen this way as well.

\subsection*{Acknowledgement}
The author thanks Nicolai Friis for discussions.
Support from the ERC CoG QITBOX, the  Generalitat  de  Catalunya  (SGR
875), Fundacion Cellex,  and Spanish MINECO (FOQUS FIS2013-46768-P and SEV-2015-0522) is acknowledged.

\end{document}